\documentclass[12pt]{article}

\usepackage{amsmath}
\usepackage{amssymb}

\begin{document}

\title{Nonconservation of the Quantum Number $K$ and Phase Transitions in Rapidly
Rotating Nuclei
\footnote{Yad. Fiz. {\bf 63} (3), 440--443 (2000) [Phys. Atomic Nuclei,
{\bf 63,} No.~3, 373--376 (2000)]}}

\author{ A.M. Kamchatnov$^{\dagger }$and V.G. Nosov$^{\ddagger}$\\
$^{\dagger}${\small\it Institute of Spectroscopy, Russian Academy of Sciences,
Troitsk, Moscow Region, 142190 Russia}\\
$^{\ddagger}${\small\it Russian Research Center Kurchatov Institute, pl. Kurchatova 1,
Moscow, 123182 Russia}
}

\maketitle

\begin{abstract}
Three different effects observed in experiments with rotating nuclei---backbending,
noncollective quadrupole transitions between different levels of the same band,
and transitions that occur, in rapidly rotating nuclei, from large-$K$ isomeric
states immediately to the levels of a rotational band despite their strong
forbiddenness in $K$---are explained in terms of nonconservation of the quantum
number $K$ in such nuclei.
\end{abstract}

Shortly after the experimental discovery of backbending [1], it was assumed [2]
that this phenomenon was due to the alignment of the angular momenta of constituent
nucleons along the angular velocity vector of the nucleus. However, the physics
behind such alignment has not been clarified until now. That the above alignment
is of a collective nature is supported by the entire body of data on the levels
of rotational bands in various nuclei. In order to clarify this point in some detail,
we can consider the frequency of a rotating nucleus as a function of the angular
momentum $I$; from the graph of this function, which, at large values of $I$, has the form
\begin{equation}\label{1}
    \hbar\Omega=\frac{dE(I)}{dI},
\end{equation}
we can see that backbending is always accompanied by a fall of $\hbar\Omega$ below the
rigid-body line
\begin{equation}\label{2}
    \hbar\Omega=\frac{\hbar^2}{\mathcal{J}_0}I,
\end{equation}
where
\begin{equation}\label{3}
    \mathcal{J}_0=\frac25MR^2
\end{equation}
is the rigid-body value of the moment of inertia of the nucleus being considered, $M$ and
$R$ being its radius and mass, respectively. For the first backbending, such a correlation
is observed for each rotational band of a nonspherical nucleus. In contrast to this,
secondary backbends occur completely under the rigid-body line, whereas so-called
low-frequency anomalies [3,4] show no correlation with this line. By way of illustration,
the rotational frequency as a function of $I$ is plotted in Fig.~1 for three rotational
bands of the $^{160}$Yb nucleus (experimental data on the transition energies for this nucleus
were taken from [5]). For each of these bands, we see that, in the region of the first
backbending, the rotational frequency as a function of $I$ traverses the rigid-body line
in the downward direction. In the rotational band built on the ground state of the $^{160}$Y
nucleus (that is, in the yrast line of this nucleus), the second backbending occurs below
the rigid-body line. In the rotational band built on the $I^{\pi} = 6^-$ state, the
low-$I$ irregularity of $\Omega$ represents a low-frequency anomaly. Such anomalies are
probably due to a noncollective alignment of an ``odd" angular momentum, which is weakly
coupled to the nuclear core [6,7].

The above feature of the experimental $I$ dependencies of $\Omega$ suggests [8] that the
first backbendings have a specific collective origin that distinguishes them both from
higher backbendings and from low-frequency anomalies. Experimental data indicate that
backbending has nothing to do with intersections of different bands; therefore, it may
be viewed as an intrinsic property of a given band. In [8], we hypothesized that
backbending results from nonconservation on the quantum number $K$, the projection of
the total angular momentum onto the symmetry axis of the nucleus. Generally, the quantity
$K = \mathbf{I}\cdot \mathbf{n}$ is well defined only in the limit $\Omega\to 0$, that is
for a nucleus at rest. As soon as components with different values of $K$ appear in the
rotational density matrix of a nucleus, the angle $\theta$ between the direction of angular
momentum $\mathbf{I}$ and the symmetry axis $\mathbf{n}$ of the nucleus becomes uncertain.
For the spin of the nucleus in excess of some critical value, $I_c$, all $K$ in the interval
$-I < K < I$ become equiprobable. In this way, a smooth evolution of the scheme governing
the coupling of angular momenta in a rotating nucleus is completed, resulting in a simple
form of this coupling; the most typical mean values are then given by
\begin{equation}\label{4}
    \langle K\rangle=0,\quad \langle K^2\rangle=\frac{I(I+1)}3,\quad
    \langle\cos^2\theta\rangle=\frac13;
\end{equation}
that is, we go over from a less symmetric to a more symmetric rotational state:
in particular, the distribution of the angle $\theta$ becomes isotropic [see the last relation
in (4)]. The critical value of the nuclear spin, $I_c$, corresponds to the first
backbending of the rotational band being considered; experimentally, this backbending
manifests itself as the above intersection of the rigid-body line.

The above collective (macroscopic) transition that involves a change in symmetry is
naturally formulated within the theory of phase transitions due to Landau [9].
In [8,10-12], the required formalism was developed in terms of the order parameter
\begin{equation}\label{5}
    \eta=1-\frac{3\langle K^2\rangle}{I(I+1)}.
\end{equation}
which vanishes in the more symmetric $I\geq I_c$ phase---we refer to it as the
$\mathbf{n}$ phase. At the point where $\Omega(I)$ intersects the rigid-body line in
the region of the first backbending, we have
\begin{equation}\label{6}
    \hbar\Omega_{nc}=\frac{\hbar^2}{\mathcal{J}_0}I_c,
\end{equation}
where $\Omega_{nc}$ is the value of the rotational frequency at $I$ values just above
$I_c$---that is, at the onset of the $\mathbf{n}$ phase.

Let us describe the situation in some detail. In the limit of adiabatically slow
rotation, the wave function can be represented as
$$
D_{KM}^I(\mathbf{n})\chi_K(\xi).
$$
If, however, the values of $I$ and $M = I_z$ are fixed, the total wave function of a
nonspherical nucleus generally has the form
\begin{equation}\label{7}
    \Psi_{IM}=\sum_{K=-I}^I D_{KM}^I(\mathbf{n})c_K^I\chi_K(\xi).
\end{equation}
In order to obtain a self-consistent description in terms
of the collective variable $\mathbf{n}$, we assume, as usual, that
the extrinsic and intrinsic values of $K$ coincide:
\begin{equation}\label{8}
    \mathbf{I}\cdot \mathbf{n}\equiv K_{\xi}.
\end{equation}
Upon the convolution of the internal (non-rotational) variables $\xi$, the relevant
density matrix takes the form
\begin{equation}\label{9}
    \begin{split}
    \rho(\mathbf{n},\mathbf{n'})&=\sum_{K=-I}^Iw(K) D_{KM}^I(\mathbf{n})
    D_{KM}^{I*}(\mathbf{n'}),\\
    w(K)&=|c_K^I|^2.
    \end{split}
\end{equation}
In the representation of three angular-momentum variables $I$, $M$, and $K$, the
rotational density matrix has the particularly simple form
\begin{equation}\label{10}
    \rho_{I_0M_0}(I,M,K;I',M',K')=\delta_{II_0}\delta_{I'I_0}\delta_{MM_0}
    \delta_{M'M_0}\delta_{KK'}w(K).
\end{equation}
Generally, the rotational state $\delta_{KK'}w(K)$ is mixed---it
becomes pure and factorizes only for the degenerate case where $K$ assumes a definite value.
In the region $I\geq I_c$, where the scheme of coupling of angular momenta is simplified,
we have $w(K) = (2I + l)^{-1}$.

The proposed interpretation of the backbending phenomenon must be verified or disproved
by confronting other predictions of the underlying approach with data. One of the
possible checks was indicated in [8]. The probability $w$ of electric quadrupole transitions
between different levels of a band is known to depend on $K^2$. Therefore, a calculation of
the lifetime of a state that picks up contributions characterized by different values of
$K$ necessarily involves averaging over the $K$
distribution. In the semiclassical approximation, the probability $w$ for $I\gg 1$ was
derived in [8,12] as
\begin{equation}\label{11}
    w=\frac{e^2\omega^5}{360\hbar c^5}\left(1+\frac{\eta}2\right)^2Q_0^2,
\end{equation}
where $\omega=2\Omega$ and $Q_0$ is the macroscopic quadrupole moment of the nucleus
due to its deformation. At the same time, the result at $K = 0$ is
$$
w=\frac{e^2\omega^5}{40\hbar c^5}\frac{I(I-1)}{(2I-1)(2I+1)}Q_0^2.
$$
By somewhat formally defining the quantity $Q_I$ as
\begin{equation}\label{12}
    Q_I^2=\frac49\left(1+\frac{\eta}2\right)^2Q_0^2,
\end{equation}
we finally arrive at the interpolation formula
\begin{equation}\label{13}
    w=\frac{e^2\omega^5}{40\hbar c^5}\frac{I(I-1)}{(2I-1)(2I+1)}Q_I^2,
\end{equation}
which holds approximately for all values of $I$. We can see that, in the $\mathbf{n}$
phase ($I\geq I_c$), the effective quadrupole moment for radiative transitions,
$Q_I$ amounts to only two-thirds of the quadrupole moment $Q_0$ for the lowest value of
$I$ in the yrast band of an even-even nucleus. The predictions are conveniently compared
with the data in terms of the ratio $Q_I^2/Q_0^2$, where $Q_0$ is the quadrupole moment
for the $2^+ \to 0^+$ transition. We have
\begin{equation}\label{14}
    \zeta=\frac{Q_I^2}{Q_0^2}=\frac2{15}\frac{(2I-1)(2I+1)}{I(I-1)}
    \left(\frac{\omega_{2\to 0}}{\omega_{I\to I-2}}\right)^5
    \frac{w_{I\to I-2}}{w_{2\to 0}}.
\end{equation}

That the probability of quadrupole transitions in a rotational band decreases with
increasing $I$ was indeed observed experimentally (see, for example, [13]), but no
attempt has been made to relate this phenomenon, referred to as the loss of collectivity,
to nonconservation of $K$ in rapidly rotating nuclei. The observed $I$ dependence of the
ratio $\zeta_I$ for the yrast line of the $^{160}$Yb nucleus [14] is displayed in Fig.~2
along with our predictions. As a rule, $\zeta_I\leq 1$, and the experimental values of
$\zeta_I$ tend to decrease with increasing $I$. Because of large uncertainties, the
agreement with the prediction is not compelling.

There was one more experimental finding that supports the hypothesis of $K$
nonconservation in rapidly rotating nuclei. We mean here the observation of direct
radiative transitions from the large-$K$ isomeric states of the $^{182}$Os, $^{174}$Hf,
and $^{179}$W nuclides to the levels of the rotational bands built on the corresponding
ground states [15-19]. Had $K$ been constant throughout the above bands, such radiative
transitions would be strongly forbidden. That these transitions occur only to higher
levels in the vicinity of backbending suggests that the amount of $K$ nonconservation
increases as $I$ approaches the critical spin value $I_c$. Yet another important observation
is that some transitions to different bands of the $^{174}$Hf nucleus (namely, to the
ground-state band, to the band built on the $\beta$-vibrational state of the nucleus,
and to the octupole and hexadecapole bands) are characterized by very similar delay factors
$f_{\nu}$ ($\nu = |K_i - K_f| - \lambda$ is the order of forbiddenness in $K$, $\lambda$
being the multipolarity of the transition from the state $i$ to the state $f$). This
implies that violation of the quantum number $K$ has a universal character independent of
the structure of a particular band. Likewise, the data are consistent with the assumption
that final states are broadly spread in $K$.

To summarize, nonconservation of the quantum number $K$ provides a clue to understanding
the aforementioned phenomena. Experimental data suggest that $K$ is violated because the
scheme of coupling of angular momenta gradually evolves as $I$ increases within a band.

\bigskip

\bigskip

\centerline{\bf Figures captions}

\bigskip

Fig.~1. Rotational frequency $\Omega$ as a function of the angular momentum $I$ (a) for the
yrast line of the $^{160}$Yb nucleus and (b) for the $I^{\pi}=6^-$ (closed circles)
and $9^-$ (open circles) bands of the same nucleus. For the yrast line of $^{160}$Yb,
the critical value of the angular momentum is 12.5. The straight line corresponds to
the rigid-body approximation.

\bigskip

Fig.~2. Squared ratio of the effective quadrupole momentum in the yrast line of
$^{160}$Yb to the quadrupole momentum for the $2^+\to 0^+$ transition [see equation (14)]
versus the angular momentum $I$.

\end{document}